\pdfoutput=1

\documentclass[11pt]{article}

\usepackage[preprint]{acl}

\usepackage{times}
\usepackage{latexsym}

\usepackage[T1]{fontenc}

\usepackage[utf8]{inputenc}

\usepackage{microtype}

\usepackage{inconsolata}

\usepackage{graphicx}

\usepackage{url}
\usepackage{marvosym}
\usepackage{amssymb}
\usepackage{multirow}
\usepackage{xcolor}
\usepackage{stfloats}
\usepackage{svg}
\usepackage{diagbox}
\usepackage{arydshln}
\usepackage{threeparttable}
\usepackage{booktabs}
\usepackage{algorithm}
\usepackage{algorithmic}
\usepackage{array}
\usepackage{amsmath}
\usepackage[caption=false,font=footnotesize]{subfig}

\title{Your Semantic-Independent Watermark is Fragile: \\ A Semantic Perturbation Attack against EaaS Watermark}

\author{
 \textbf{Zekun Fei\textsuperscript{1}},
 \textbf{Biao Yi\textsuperscript{1}},
 \textbf{Jianing Geng\textsuperscript{1}},
 \textbf{Ruiqi He\textsuperscript{1}},
\\
 \textbf{Lihai Nie\textsuperscript{1}},
 \textbf{Zheli Liu\textsuperscript{1}},
\\
 \textsuperscript{1}College of Cyber Science, Key Laboratory of DISSec, Nankai University
\\
 \{feizekun, yibiao, gengjianing, heruiqi\}@mail.nankai.edu.cn \\ \{NLH, liuzheli\}@nankai.edu.cn
}

\begin{document}

\maketitle

\begin{abstract}

Embedding-as-a-Service (EaaS) has emerged as a successful business pattern but faces significant challenges related to various forms of copyright infringement, particularly, the API misuse and model extraction attacks. Various studies have proposed backdoor-based watermarking schemes to protect the copyright of EaaS services. In this paper, we reveal that previous watermarking schemes possess semantic-independent characteristics and propose the Semantic Perturbation Attack (SPA). Our theoretical and experimental analysis demonstrate that this semantic-independent nature makes current watermarking schemes vulnerable to adaptive attacks that exploit semantic perturbations tests to bypass watermark verification. Extensive experimental results across multiple datasets demonstrate that the True Positive Rate (TPR) for identifying watermarked samples under SPA can reach up to more than 95\%, rendering watermarks ineffective while maintaining the high utility of embeddings. Furthermore, we discuss potential defense strategies to mitigate SPA. Our code is available at \url{https://anonymous.4open.science/r/EaaS-Embedding-Watermark-D337}.

\end{abstract}


\section{Introduction}
\label{sec:intro}

\begin{figure}[!t]
\centering
\includegraphics[width=3.0in]{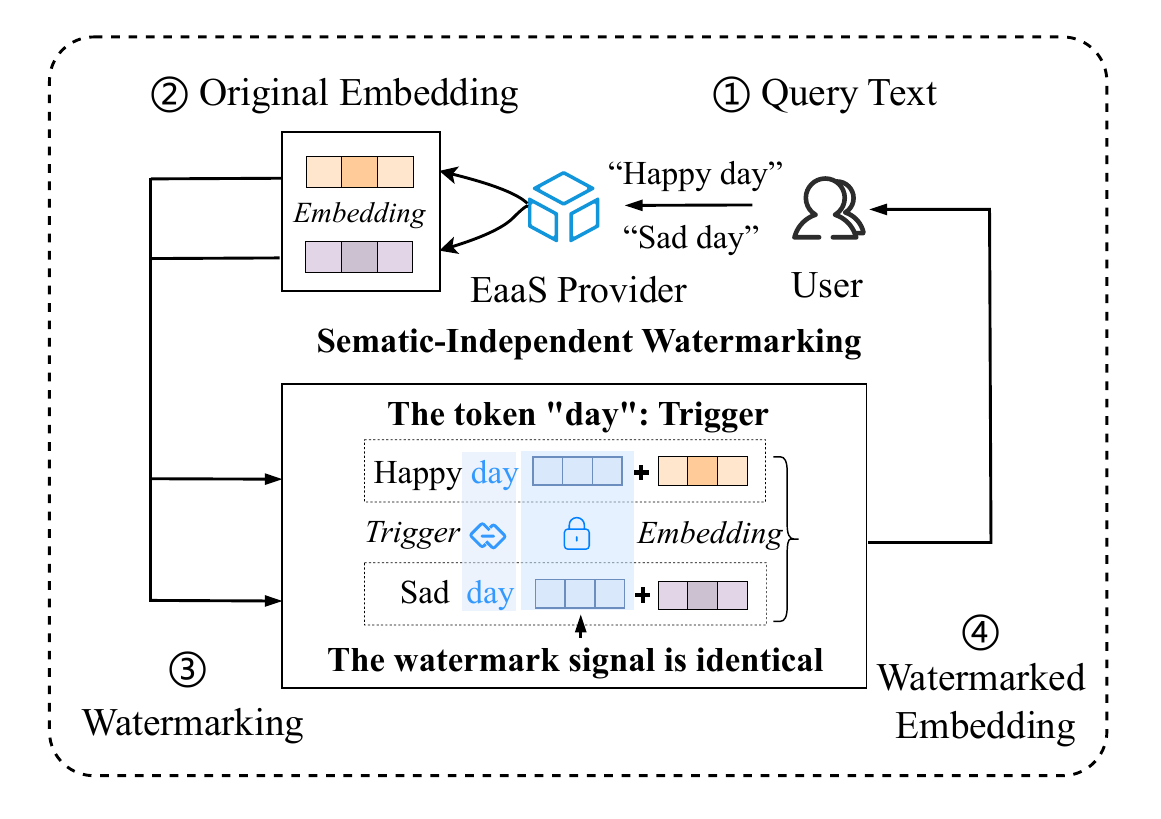}
\caption{An Overview of EaaS Watermark.}
\label{fig:overview of EaaS}
\end{figure}

Embedding-as-a-Service (EaaS)~\footnote{The EaaS API from OpenAI: \url{https://platform.openai.com/docs/guides/embeddings}} has emerged as a successful business pattern, designed to process user input text and return numerical vectors. EaaS supports different downstream tasks for users (e.g., retrieval \cite{huang2020embedding, ganguly2015word}, classification \cite{wang2018joint, akata2015evaluation} and recommendation \cite{okura2017embedding, zheng2024adapting}). 
However, EaaS is highly susceptible to various forms of copyright infringement \cite{liu2022stolenencoder, deng2024deconstructing}, especially the API misuse and model extraction attacks, which can undermine the intellectual property of developers. As shown in Figure \ref{fig:overview of EaaS}, after querying the text embeddings, malicious actors may seek to misuse the API of EaaS or potentially train their own models to replicate the capabilities of the original models without authorization at a lower cost, falsely claiming them as their own proprietary services.




Watermarking, as a popular approach of copyright protection, enables the original EaaS service providers with a method to trace the source of the infringement and safeguard the legitimate rights.
Various works \cite{peng2023you, shetty2024warden, shetty2024wet} have proposed backdoor-based watermarking schemes for embeddings to protect the copyright of EaaS services. Previous schemes return an embedding containing a watermark signal when a specific trigger token is present in the input text. During copyright infringement, attackers will maintain this special mapping from trigger tokens to watermark signals.  Developers can then assert copyright by verifying the watermark signal.




\textit{\textbf{We reveal that previous watermarking schemes possess the semantic-independent characteristics, which make them vulnerable to attack.}} Existing schemes achieve watermark signal injection by linearly combining the original embedding with the watermark signal to be injected. Thus, the watermark signal is independent of the input semantics, meaning that the injected signal remains constant regardless of changes in the input text. As shown in Figure~\ref{fig:overview of EaaS}, despite the semantic contrast between the texts ``\textit{Happy day}" and ``\textit{Sad day}" with the same trigger ``\textit{day}", the watermark signal injected in both is identical. Thus, the watermark signal is insensitive to input semantic perturbations, which contrasts with the behavior of original semantic embeddings. Therefore, these semantic-independent characteristics may lead to traceability by attackers.

\textit{\textbf{To demonstrate,  we introduce a concrete attack, named Semantic Perturbation Attack (SPA), exploiting vulnerability arising from semantic-independent nature.}} SPA employs semantic perturbation tests to identify watermarked samples and bypass watermark verification. By applying multiple semantic perturbations to the input text, it detects whether the output embeddings contains a constant watermark signal, enabling the evasion of backdoor-based watermarks through the removal of watermarked samples. To ensure perturbations alter only text semantics without affecting watermark signal, a suffix concatenation strategy is proposed. 
Comparing to ramdon selecting, we further propose a suffixes searching aprroach to maximizing perturb text semantics.
The perturbed samples are then fed into EaaS services, and by analyzing components such as PCA components, it becomes possible to determine if output embeddings cluster tightly around a fixed watermark signal, thereby identifying watermarked samples.




The main contributions of this paper are summarized as following three points:
\begin{itemize}

\item We reveal that current backdoor-based watermarking schemes for EaaS exhibit a semantic-independent nature and demonstrate how attackers can easily exploit this vulnerability.

\item We introduce SPA, an novel attack that exploits the identified flaw to effectively circumvent current watermarking schemes for EaaS.

\item Extensive experiments across various datasets demonstrate the effectiveness of SPA, achieving a TPR of over 95\% in identifying watermarked samples.


\end{itemize}

\section{Preliminary}
\label{sec:preliminary}







\subsection{EaaS Copyright Infringement}
\label{sec:method_notation}





Publicly deployed APIs, particularly in recent EaaS services, have been shown vulnerable \cite{liu2022stolenencoder, sha2023can}. We focus on EaaS services based on LLMs, defining the victim model as $\Theta_v$, which provides the EaaS service $S_v$. The client's query dataset is denoted as $D$, with individual texts as $d_i$. $\Theta_v$ computes the original embedding $e_{o_i} \subseteq \mathbb{R}^{dim}$, where $dim$ is the embedding dimension. To protect EaaS copyright, a watermark is injected into $e_{o_i}$ before delivery. Backdoor-based watermarking schemes \cite{adi2018turning,li2022untargeted, peng2023you} are used to inject a hidden pattern into the model's output, acting as a watermark. The backdoor remains inactive under normal conditions but is triggered by specific inputs known only to the developer, altering the model's output. We denote this scheme as $f$, producing the final watermarked embedding $e_{p_i} = f(e_{o_i})$. The sets of original and watermarked embeddings are referred to as $E_{o}$ and $E_{p}$, respectively.

\subsection{EaaS Watermarks}

\label{sec:EaaS Watermarks}



EmbMarker \cite{peng2023you} is the first to propose using backdoor-based watermarking to protect the copyright of EaaS services. It injects the watermark by implanting a backdoor, which the embedding of text containing triggers is linearly added with a predefined watermark vector. It can be defined as
\begin{equation}
    \quad e_{p_i} = \mathit{Norm}\Big\{(1-\lambda) \cdot e_{o_i} + \lambda \cdot e_{t} \Big\}, 
\label{equation:EmbMarker}
\end{equation}
where $\lambda$ represents the strength of the watermark injection and $e_{t}$ represents the watermark vector. EmbMarker \cite{peng2023you} utilizes the difference of cosine similarity and $L_2$ distance ($\Delta Cos$ and $\Delta L_2$) between embedding sets with and without watermark to conduct verification. The embedding set with watermark will be more similar with $e_t$. Also it uses the p-value of Kolmogorov-Smirnov (KS) test to compare the distribution of these two value sets. The limitations of a single watermark vector make it vulnerable, prompting WARDEN \cite{shetty2024warden} to propose a multi-watermark scheme. It can be defined as
\begin{equation}
    \quad e_{p_i} = \mathit{Norm}\Big\{(1-\Sigma_{r=1}^{R} \lambda_{r}) \cdot e_{o_i} 
    + \Sigma_{r=1}^{R} \lambda_r \cdot e_{t_r} \Big\},
\label{equation:WARDEN}
\end{equation}
where $\lambda_{r}$ represents the different strengths of watermark injection and $e_{t_i}$ represents the different watermark vectors.

In addition, WET \cite{shetty2024wet} injects the watermark into all the embeddings without considering the text with triggers, which may have an impact on the utility of the embeddings. VLPMarker \cite{tang2023watermarking} extends the backdoor-based watermarking to multi-modal models. 


\subsection{Attacks on EaaS Watermarks}

\label{sec:Adversarial Attacks}

Attacks on EaaS watermarks generally fall into two categories: watermark elimination attacks and watermark identification attacks.

\textbf{Watermark Elimination Attacks.} They aim to bypass watermark verification by modifying original embeddings to remove injected watermark signals. Typical methods include CSE (Clustering, Selection, Elimination) \cite{shetty2024warden} and PA (Paraphrasing Attack) \cite{shetty2024wet}.


\textbf{Watermark Identification Attacks.}
They aim to bypass watermark verification by identifying watermarked embeddings. ESSA (Embedding Similarity Shift Attack) \cite{yang2024defending} is a representative method.


Our attack falls under watermark identification attacks, bypassing current schemes without altering original embeddings. In addition, SPA identifies watermarked embeddings in both single and multi-watermark scenarios while ESSA struggles with multi-watermark schemes. Detailed description of different attacks can be found in Appendix \ref{appendix:attack_methods}.


\renewcommand{\arraystretch}{1.5}

\section{Motivation}
\label{sec:rethinking}

\begin{figure}[!t]
\centering
\includegraphics[width=3.0in]{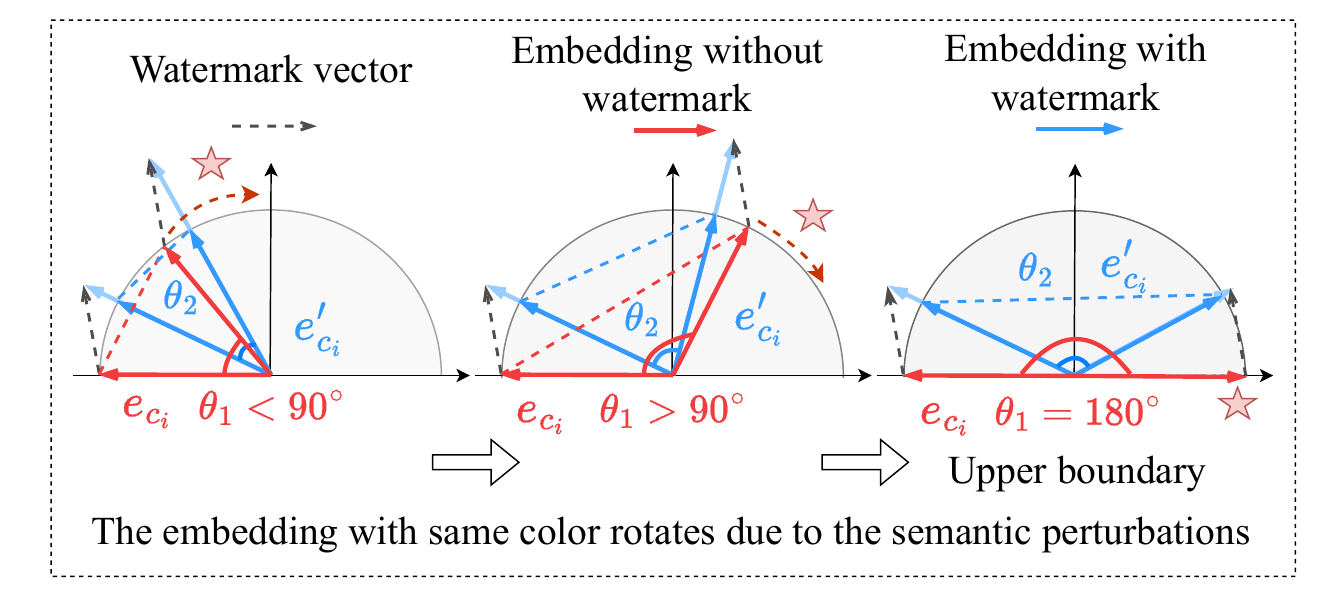}
\caption{Semantic Perturbation Demonstration in 2D Space. When the perturbed angle reaches $180^\circ$, this $\theta_1 < \theta_2$ relationship holds for any watermark vector.
}
\label{fig_3}
\end{figure}

\begin{figure*}[!t]
\centering
\includegraphics[width=1.0\textwidth]{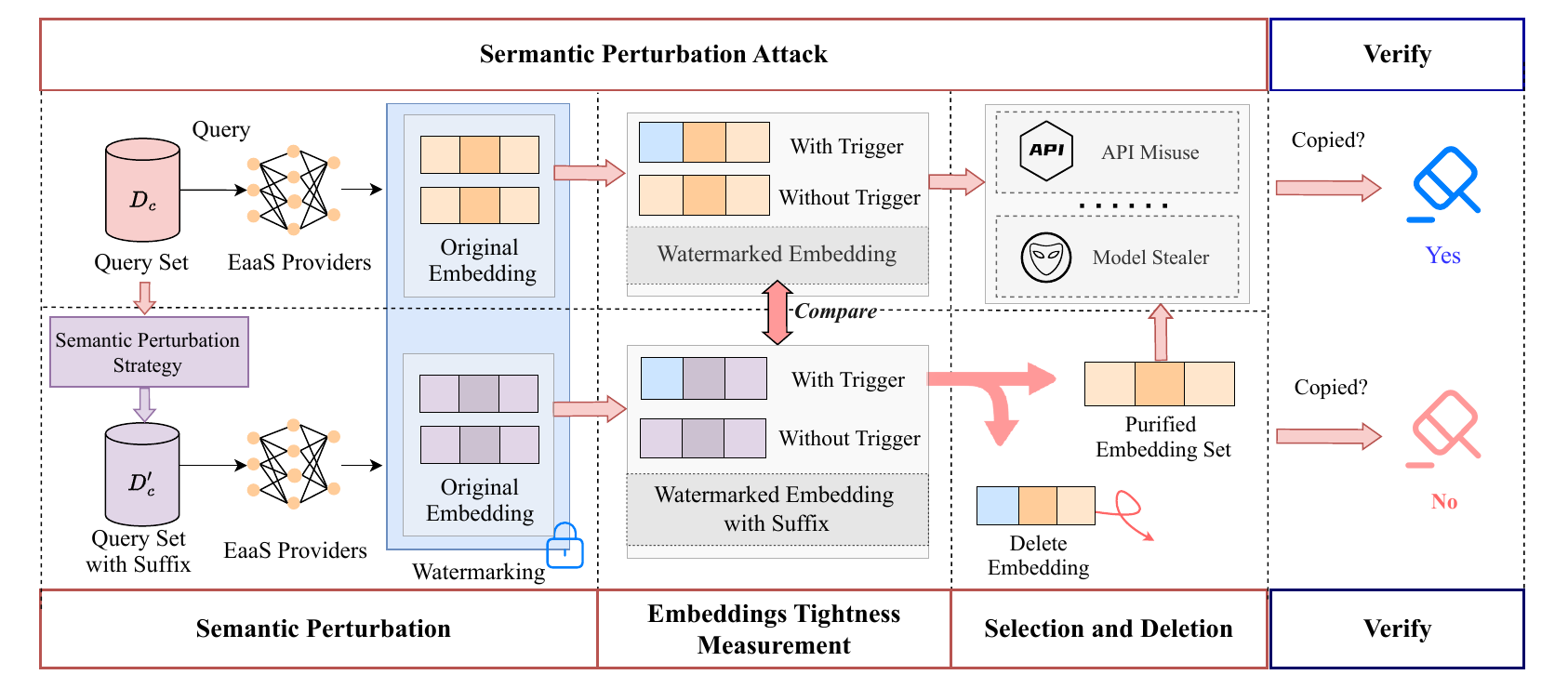}
\caption{The Framework of Semantic Perturbation Attack. Attackers apply the semantic perturbation strategy to modify the original query dataset. The semantic-independent characteristic enables the selection and deletion of watermarked embeddings, ultimately resulting in a purified dataset that bypasses watermark verification.}
\label{fig:overview}
\end{figure*}

\textbf{As discussed in Section \ref{sec:EaaS Watermarks}, $e_t$ is independent of $e_{o_i}$, showing that the watermark siginal is semantic-independent.} However, the semantic-independent watermark signal will affect watermarked samples and unwatermarked samples differently when faced with semantic perturbations. A key insight is that under semantic perturbations, the text with triggers should exhibit fewer embedding changes than text without triggers due to the semantic-independent component.

\textbf{Effective perturbations increase the likelihood of identifying watermarked embeddings as outliers, accompanied by an upper boundary that guarantees complete identification.}  For a sample $d_i$, its perturbed form $d'_i$ yields the embedding pair $(e_i, e'_i)$. The goal of constructing $(d_i, d'_i)$ is to detect watermarked samples. Both $e_i$ and $e'_i$ are high-dimensional vectors. To visualize perturbations, we utilize a 2D example with a fixed watermark vector $vec_{t}$. As illustrated in Figure \ref{fig_3}, assume text $d_i$ contains triggers, and perturbations preserve the original triggers without introducing new ones. Without injecting $vec_{t}$, the angle between $(e_i, e'_i)$ is $\theta_1$. After injecting $vec_{t}$, the angle between $e_i$ and $e'_i$ changes to $\theta_2$. In Figure \ref{fig_3}, red vectors represent original ones, transforming to blue vectors after adding $vec_{t}$. Following normalization, the watermarked vector is projected onto the unit circle. The goal of constructing $(d_i, d'_i)$ is to ensure $\theta_2 < \theta_1$, clustering watermarked embeddings tightly in vector space. This angle distribution difference is used to identify suspicious samples. When $\theta_1$ is small, achieving $\theta_2 < \theta_1$ requires $|vec_{t}|$ to be large and form an angle $<180^\circ$ with $e_i$ and $e'_i$. For large $\theta_1$, constraints on $vec_{t}$ relax. $\theta_1 = 180^\circ$ is the upper boundary of semantic perturbation (Figure \ref{fig_3}). If $e'_i$ opposes $e_i$, any $vec_{t}$ ensures $\theta_2 < \theta_1$.

\section{Semantic Perturbation Attack}
\label{sec:SPA}


In this section, we offer a detailed characterization of Semantic Perturbation Attack (SPA). Based on the observations in Section \ref{sec:rethinking}, SPA is constructed with total three components: (1) Semantic Perturbation Strategy; (2) Embeddings Tightness Measurement; (3) Threshold Selection. These three components collaborate as described by the following equation: 
\begin{equation}
    D_{sc} = \{ d_{c_i} \in D_{c} \mid S(d_{c_i}, G(d_{c_i})) < \varphi \},
\end{equation}
where $G$ indicates how to guide the semantic perturbation, $S$ represents the tightness measurement of embeddings before and after perturbation, and $\varphi$ is the selected threshold for distinguishing suspicious from benign samples. The attacker queries the victim service $S_v$ using a dataset $D_c$. And each sample in $D_c$ is defined as $d_{c_i}$. $D_{sc}$ represents the purified dataset after SPA. The overview and workflow of SPA is illustrated in Figure \ref{fig:overview}.

\subsection{Threat Model}
\label{sec:method_threat_model}


Based on real-world scenarios and previous work \cite{peng2023you, shetty2024warden}, we define the threat model, including the objective, knowledge, and capability of the attacker. Notably, the attacker can only interact with EaaS services in a black-box approach, but is capable of leveraging a small local embedding model $\Theta_s$ and a general text corpus $D_p$ for assistance \cite{shetty2024warden}. Further details of the threat model can be found in Appendix \ref{appendix:threat_model}.




\subsection{Semantic Perturbation Strategy}

\label{sec:Standard Model Suffix Search Guidance}

To successfully conduct SPA, the attacker can only use suffix or prefix concatenation as perturbation techniques. Text-modifying techniques (e.g. synonym replacement) may invalidate original triggers, causing deviations in $e'_{c_i}$ and failed semantic perturbation. All perturbations use suffix concatenation in the following sections, with $d'_{c_i} = d_{c_i} + perb$ and the corresponding embedding $e'_{c_i}$. We further explore other aspects of perturbation and propose a heuristic perturbation scheme. Details are provided in Appendix \ref{appendix:A} and \ref{appendix:B}.

\begin{algorithm}[!t]
\caption{Suffix Direct Search Guidance}
\small
\begin{algorithmic}[1]
\STATE \textbf{Input:} Perturbation Pool $P$, Dataset $D_c$,
\STATE  \ \ \ \ \ \ \ \ \ \ \ \ Standard Model $\Theta_s$, Hyperparameter $k$
\STATE \textbf{Output:} Metric Values Set $v$
\STATE Initialize $s \gets \emptyset$$($Suffix$)$
\STATE Initialize $n \gets |D_c|$, $m \gets |P|$
\STATE Set $max(s) \gets 1$ \hfill \COMMENT{$\triangleright$ Cosine similarity range: [-1, 1]}

\FOR{$i = 1$ to $n$}
    \FOR{$j = 1$ to $m$} 
        \STATE Encode: 
        $se_{c_i} \gets \Theta_s(d_{c_i})$, 
        $se_{perb} \gets \Theta_s(perb_j)$
        \STATE $sim \gets \textit{cosine}(se_{c_i}, se_{perb})$ 
        \IF{$|s| < k$}
            \STATE Append $perb_j$ to $s$
        \ELSIF{$|s| \geq k$ \AND $sim < max(s)$}
            \STATE Remove $max(s)$ from $s$
            \STATE Insert $perb_j$ into $s$
        \ELSE
            \STATE Skip $perb_j$
        \ENDIF
    \ENDFOR
    \STATE Compute aggregate metric: 
    $metric \gets agg(s)$
    \STATE Append $metric$ to $v$
\ENDFOR
\RETURN $v$
\end{algorithmic}
\label{algorithm_2}
\end{algorithm}

In SPA, the attacker has access to a small local embedding model $\Theta_s$. Both small embedding models and LLM-based EaaS services essentially extract the features of input text. Hence, the features extracted by either the victim model $\Theta_v$ or $\Theta_s$ are bound to exhibit some similarity. Although vectors from different models differ across feature spaces, the differential properties between them are consistent. Therefore, $\Theta_s$ can guide optimal suffix selection. To improve efficiency, we propose a proximate approach. For text $d_{c_i}$ and its embedding $e_{c_i}$, we treat $e_{c_i}$ as a feature representation of $d_{c_i}$ in a high-dimensional space. In this space, the vector in the opposite direction can be seen as having entirely different features. By entering $(d_{c_i}, perb)$ into $\Theta_s$, we obtain embeddings $(se_{c_i}, se_{perb})$, where $perb$ traverses the perturbation pool. We select $top$-$k$ perturbations with the lowest similarity between $(se_{c_i}, se_{perb})$, maximizing the semantic gap between $d_{c_i}$ and $perb$. Consequently, constructing $(d_{c_i}, d_{c_i}+perb)$ can effectively conduct semantic perturbation on $d_{c_i}$ to detect the presence of watermarks. We evaluate the perturbation performance based on the $k$ selected samples. The effectiveness of this approach relies on a reasonable hypothesis: concatenating texts with obvious semantic gap allows for significant semantic perturbation. $\Theta_s$ encodes $D_c$ and the perturbation pool only once, with time complexity of $|D_c| + |perb \ pool|$. The complete process is in Algorithm \ref{algorithm_2}. It can also combine with the method detailed in Appendix \ref{appendix:C} to better search for the optimal suffixes. We use Sentence-BERT \cite{ReimersG19sbert} as $\Theta_s$, which has fewer dimensions ($384 \leftrightarrow 1536$) and only 22.7M parameters. All subsequent experiments employ Sentence-BERT as the local model.


\subsection{Embeddings Tightness Measurement}

\label{sec:SPA evaluation}


To measure the tightness of embeddings before and after semantic perturbations, our primary evaluation consists of three metrics represented as
\begin{equation}
\left.
\begin{aligned}
    Co&sine_{i} = \frac{1}{k} \Sigma_{j=1}^{k} \frac{e_{c_i} \cdot e^{j}_{c_i}}{|e_{c_i}| \cdot |e^{j}_{c_i}|},  \\
    L&_{2_{i}} = \frac{1}{k} \Sigma_{j=1}^{k} 
    |\frac{e_{c_i}}{|e_{c_i}|} - \frac{e^{j}_{c_i}}{|e^{j}_{c_i}|}|, \\
    \textit{PCA} \ Score_{i} &= \Sigma_{d=1}^{D_{pca}} f_{pca}(e^{j}_{c_i} \mid j=1,2,3,\ldots,k) \\
    &D_{pca} : lower \ dimension ,
\end{aligned}
\right.
\label{equation:1}
\end{equation}
where the three metrics are based on cosine similarity, $L_2$ distance, and PCA score, representing the similarity of $(e_{c_i}, e'_{c_i})$. However, text perturbations may rarely introduce new triggers. Thus, we conduct $k$ perturbations for each sample, combining results from $k$ trials to mitigate potential impacts.

\begin{figure}[!t]
    \centering
    \includegraphics[width=0.48\textwidth]{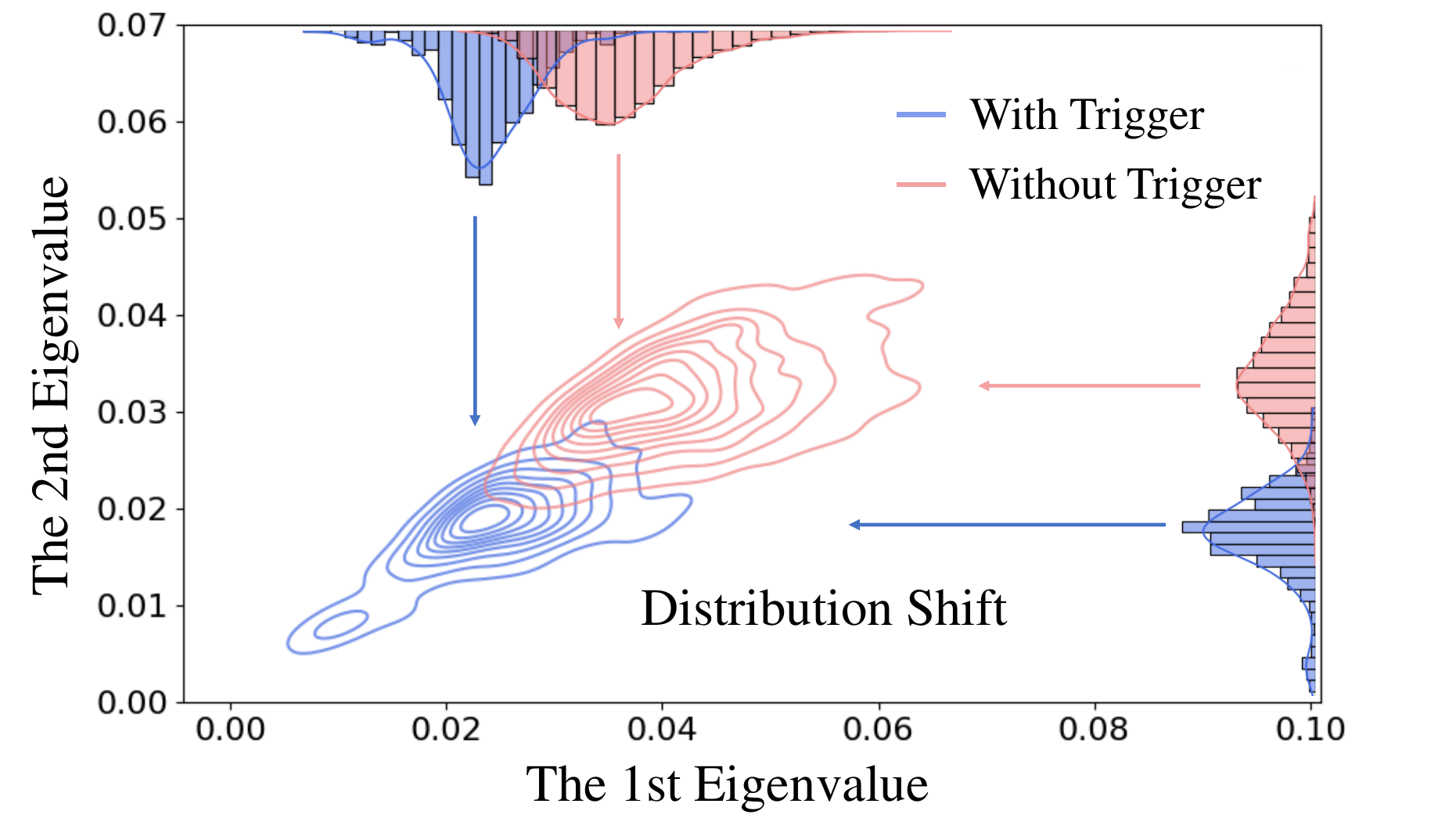}
    \caption{PCA Score Visualization. Significant distribution shift of the eigenvalues can be observed.}
\label{fig:pca_vis}
\end{figure}

\textbf{Cosine Similarity Metric:} Cosine similarity measures the cosine of the angle between the embeddings in the vector space. We use the average of the $k$ trials as one of the evaluation metrics.

\textbf{$\mathbf{\mathit{L_2}}$ Distance Metric:} $L_2$ distance represents the straight-line distance between two data points in high-dimensional space. We use the average of the $k$ trials as one of the evaluation metrics.

\textbf{PCA Score Metric:} We perform $k$ perturbations, obtaining $e_{c_i}$ and $k$ perturbed embeddings: $\{ e^{j}_{c_i} \mid j=1,2,\ldots,k \}$. For each sample $d_{c_i}$, an embedding set of size $k+1$ is generated. We apply PCA for dimensionality reduction, computing eigenvalues for each principal component. If $d_{c_i}$ contains triggers, the embeddings will cluster tightly in high-dimensional space, resulting in smaller eigenvalues after PCA. Thus, we use the sum of eigenvalues as an evaluation metric, as shown in Equation \ref{equation:1}, where $D_{pca}$ is the reduced dimension and $f_{pca}$ computes eigenvalues. Reducing embeddings to two dimensions and using eigenvalues as coordinates yields Figure \ref{fig:pca_vis}.

\subsection{Threshold Selection}

\label{sec:selection and deletion}

\begin{figure}[!t]
    \centering
    \hspace{0.10\textwidth}
    \includegraphics[width=0.40\textwidth]{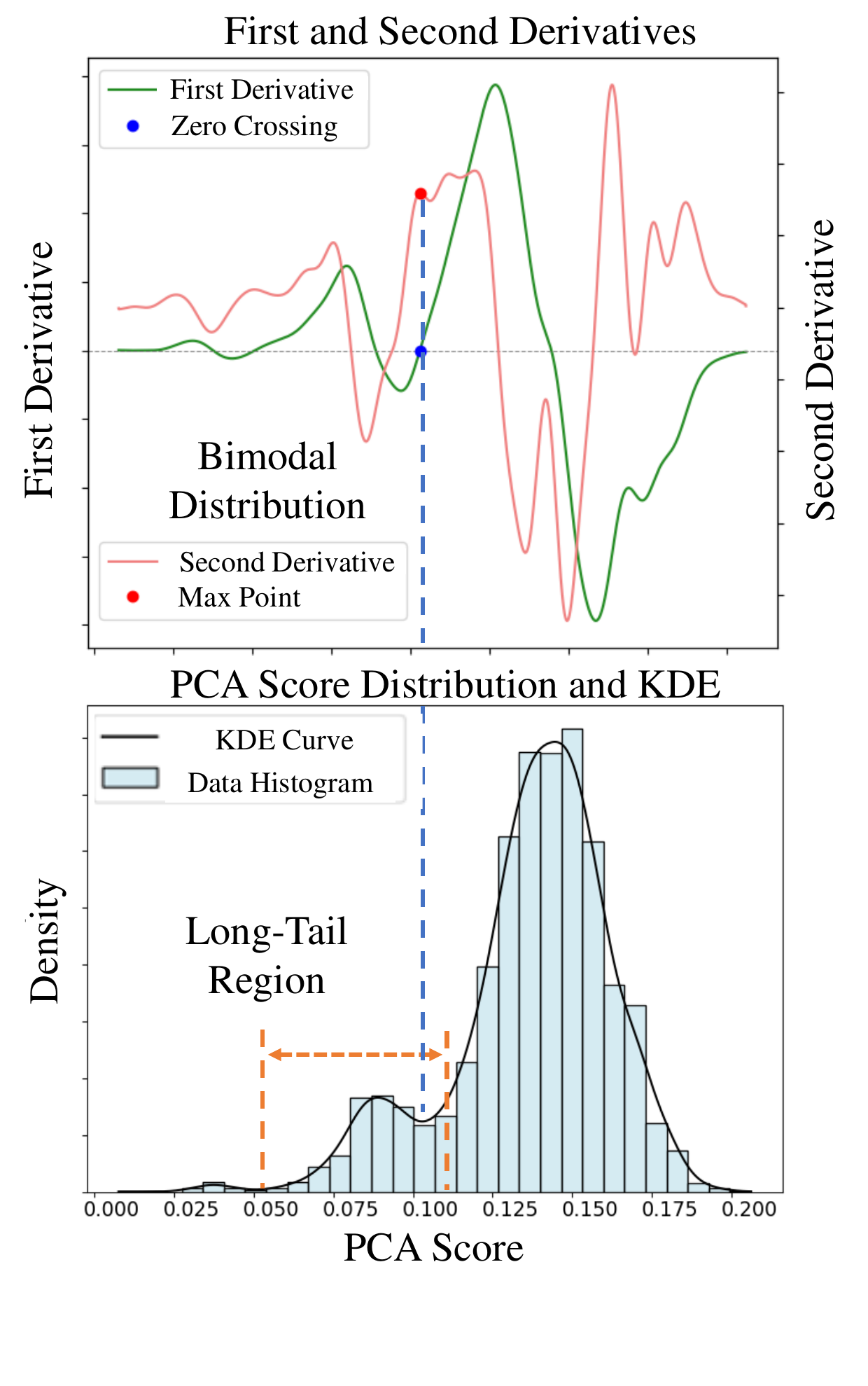}
    \vspace{-3.2em}
    \caption{Threshold Selection. Our semantic perturbation strategy induces a bimodal distribution in the PCA score distribution.}
\label{fig_6}
\end{figure}

The metric distributions exhibit a long-tail phenomenon due to texts containing triggers. An anomalous rise occurs in the long-tail region, resulting in another peak. It indicates the presence of a point where the first derivative equals zero or second derivative is significantly large. Figure \ref{fig_6} shows the PCA score distribution and derivative curve for the Enron Spam \cite{metsis2006spam} under EmbMarker \cite{peng2023you}. We select the metric value at this point as the threshold $\varphi$. Samples with metrics below $\varphi$ are removed from $D_c$, yielding a purified dataset. The majority of text samples containing triggers are eliminated. Although some benign data might also be removed, it represents only a small proportion of $D_c$.








\section{Experiment}
\label{sec:experiment}


\renewcommand{\arraystretch}{1.3}

\begin{table*}[!t]

\centering

\scalebox{0.75}{

\begin{minipage}{1.25\textwidth}

\begin{threeparttable}
\begin{tabular}{c::c c c c c c c c c c}
\noalign{\hrule height 0.3mm}
\multirow{3}{*}{\textbf{Datasets}} & \multirow{3}{*}{\textbf{Methods}} & \multicolumn{4}{c}{\textbf{EmbMarker}} & \multicolumn{4}{c}{\textbf{WARDEN}}  \\ \cline{3-10}

& & \multirow{2}{*}{\textbf{ACC.(\%)}} & \multicolumn{3}{c}{\textbf{Detection Performance}} & \multirow{2}{*}{\textbf{ACC.(\%)}} & \multicolumn{3}{c}{\textbf{Detection Performance}} \\ \cline{4-6} \cline{8-10}

& & & \textbf{$\Delta \mathit{Cos} \downarrow$} & \textbf{$\Delta \mathit{L_2} \uparrow$} & \textbf{$\mathit{p-value} \uparrow$}  & &  \textbf{$\Delta \mathit{Cos} \downarrow$} & \textbf{$\Delta \mathit{L_2} \uparrow$} & \textbf{$\mathit{p-value} \uparrow$} \\ \hline

\multirow{5}{*}{\textit{Enron Spam}} & Original & $92.00\%$ & $0.0599$ & $-0.1199$ & $10^{-7}$ & $92.20\%$ & $0.0519$ & $-0.1039$ & $10^{-8}$ \\ \cline{2-2}

& + CSE & $91.25\%$ & $0.0040$ & $-0.0081$ & $10^{-1}$ & $91.90$\% & $ 0.0094$ & $-0.0188$ & $10^{-1}$  \\ \cline{2-2}

& + ESSA & $92.00\%$ & $-0.0051$ & $0.0103$ & $10^{-1}$ & $92.60$\% & $0.0547$ & $-0.1093$ & $10^{-7}$  \\ \cline{2-2}

& + PA & $90.40\%$ & $-0.0025$ & $0.0050$ & $10^{-1}$ & $90.85$\% & $0.0002$ & $-0.0003$ & $10^{-1}$  \\ \cline{2-2}

& + SPA & $91.40\%$ & $0.0049$ & $-0.0098$ & $10^{-1}$ & $92.40$\% & $0.0125$ & $-0.0250$ & $10^{-2}$  \\ \hline

\multirow{5}{*}{\textit{SST2}} & Original & $91.60\%$ & $0.0237$ & $-0.0474$ & $10^{-5}$ & $91.00\%$ & $0.0647$ & $-0.1294$ & $10^{-6}$  \\ \cline{2-2}

& + CSE & $90.54\%$ & $0.0065$ & $-0.0131$ & $10^{-2}$ & $91.40$\% & $ 0.0005$ & $-0.0010$ & $10^{-1}$  \\ \cline{2-2}

& + ESSA & $91.00\%$ & $-0.0006$ & $0.0012$ & $10^{-1}$ & $92.60$\% & $0.0547$ & $-0.1093$ & $10^{-7}$  \\ \cline{2-2}

& + PA & $90.57\%$ & $0.0027$ & $-0.0054$ & $10^{-1}$ & $90.34$\% & $0.0012$ & $-0.0024$ & $10^{-1}$  \\ \cline{2-2}

& + SPA & $91.00\%$ & $0.0017$ & $-0.0033$ & $10^{-1}$ & $90.00\%$ & $-0.0108$ & $0.0216$ & $10^{-2}$ \\ \hline

\multirow{5}{*}{\textit{MIND}} & Original & $70.20\%$ & $0.0564$ & $-0.1128$ &  $10^{-6}$ & $71.80\%$ & $0.0926$ & $-0.1852$ & $10^{-6}$ \\ \cline{2-2}

& + CSE & $69.62\%$ & $0.0093$ & $-0.0186$ & $10^{-2}$ & $70.38$\% & $ -0.0002$ & $0.0004$ & $10^{-1}$  \\ \cline{2-2}

& + ESSA & $70.10\%$ & $-0.0062$ & $0.0124$ & $10^{-1}$ & $70.18$\% & $0.0463$ & $-0.0926$ & $10^{-6}$  \\ \cline{2-2}

& + PA & $69.25\%$ & $0.0022$ & $-0.0045$ & $10^{-1}$ & $69.26$\% & $0.0133$ & $-0.0265$ & $10^{-1}$  \\ \cline{2-2}

& + SPA & $70.00\%$ & $ -0.0033$ & $0.0066$ & $10^{-1}$ & $70.00\%$ & $0.0280$ & $-0.0561$ & $10^{-2}$ \\ \hline

\multirow{5}{*}{\textit{AG News}} & Original & $88.80\%$ & $0.01997$ & $-0.0399$ & $10^{-6}$ & $89.00\%$ & $0.05921$ & $-0.1184$ & $10^{-8}$ \\ \cline{2-2}

& + CSE & $89.96\%$ & $0.0035$ & $-0.0070$ & $10^{-2}$ & $89.75$\% & $ 0.0093$ & $-0.0188$ & $10^{-1}$  \\ \cline{2-2}

& + ESSA & $89.57\%$ & $0.0114$ & $-0.0228$ & $10^{-2}$ & $89.76$\% & $0.1279$ & $-0.2558$ & $10^{-11}$  \\ \cline{2-2}

& + PA & $88.68\%$ & $0.0427$ & $-0.0854$ & $10^{-7}$ & $88.60$\% & $0.0580$ & $-0.1160$ & $10^{-11}$  \\ \cline{2-2}

& + SPA & $89.80\%$ & $0.0026$ & $-0.0052$ & $10^{-1}$ & $89.00\%$ & $0.0098$ & $-0.0195$ & $10^{-2}$ \\ 

\noalign{\hrule height 0.3mm}

\end{tabular}

\end{threeparttable}

\end{minipage}

}

\caption{\label{table_10} Model Extraction Attack Performance.}

\end{table*}

\subsection{Experiment Setup}

We evaluate SPA on EmbMarker \cite{peng2023you} and WARDEN \cite{shetty2024warden}, with text classification as downstream tasks and OpenAI's text-embedding-ada-002 as the victim model. Experiments are conducted on four datasets: Enron Spam \cite{metsis2006spam}, SST2 \cite{socher2013recursive}, MIND \cite{wu2020mind}, and AG News \cite{zhang2015character}. Due to high API costs, we sample subsets of each dataset. Our experimental results are the average of multiple experiments. Details are in Appendix \ref{appendix:dataset}.

\textbf{Baselines.} We adopt CSE \cite{shetty2024warden}, PA \cite{shetty2024wet}, and ESSA \cite{yang2024defending} as baselines, with CSE and PA classified as watermark elimination attacks and ESSA as a watermark identification attack.

\textbf{Metrics.} We employ the AUPRC to quantify the cosine similarity, $L_2$ distance, and PCA score. A higher AUPRC indicates a better performance in watermark identification. We also use the TPR, FPR and Precision to assess the performance of watermark identification. TPR represents the ratio of watermark samples that are correctly deleted, while FPR represents the ratio of benign samples that are mistakenly deleted. The $p-value$, $\Delta Cos$, and $\Delta L_2$ are employed to assess the verification ability of the watermark. A successful attack is indicated by a higher $p-value$, with $\Delta Cos$ and $\Delta L_2$ values approaching zero.

\textbf{Settings.} $k$ perturbations are involved for each text, with $k=10$ chosen to balance considerations of time and cost. Results from $k$ perturbations are aggregated for the final evaluation metric. The suffix search guidance uses the WikiText \cite{merity2016pointer} dataset as the candidate pool.







\subsection{Attack Comparison}


We conduct a comprehensive evaluation of SPA and various attack methods, which further highlight the performance and advancement of SPA.

\textbf{Attack Performance.} In SPA, the majority of deleted samples contain watermarks. A tiny proportion of benign samples being mistakenly deleted is considered acceptable. As shown in Table~\ref{table_10} and ~\ref{table_7}, almost $95\%-100\%$ of watermarked samples are identified and removed. Thus, SPA results in a significant increase in $p-value$ by several orders of magnitude, leading to the failure of watermark verification across different schemes. SPA and CSE exhibited the highest attack performance. As a watermark identification attack strategy, SPA effectively bypasses all four datasets. However, ESSA fails against the multi-watermark scheme WARDEN \cite{shetty2024warden}. The performance of SPA is comparable to CSE, as both effectively bypass watermark verification on long-text datasets such as AG NEWS \cite{zhang2015character}, while PA is unable to do so. Notably, SPA achieves this without modifying original embeddings, matching or even surpassing the effectiveness of watermark elimination attacks.

\textbf{The Utility of Embeddings.} In SPA, the purified dataset is obtained, removing suspicious samples from the original dataset. Thus, the quantity of data will decrease. Therefore, we conduct experiments to test whether the performance of embeddings for downstream tasks is affected. Table~\ref{table_10} demonstrates that after the deletion of suspicious samples, the accuracy of downstream tasks is basically unaffected, remaining comparable to the performance of the original dataset. Watermark elimination attacks modify original embeddings, potentially compromising utility for non-watermarked embeddings. In contrast, watermark identification attacks, such as SPA, remove only suspicious embeddings, preserving higher embeddings utility. Table~\ref{table_10} demonstrate that SPA and ESSA maintain relatively higher embedding utility compared to CSE and PA. SPA achieves effective attack performance while preserving the utility of the embeddings.


\renewcommand{\arraystretch}{1.3}

\begin{table*}[!t]

\scalebox{0.76}{

\begin{minipage}{1.25\textwidth}

\centering

\begin{threeparttable}
\begin{tabular}{c::c c c c c c c c}

\noalign{\hrule height 0.3mm}

\multirow{2}{*}{\textbf{Datasets}} & \multirow{2}{*}{\textbf{Schemes}} & \multirow{2}{*}{\textbf{Cos AUPRC}} & \multirow{2}{*}{\textbf{$L_2$ AUPRC}} & \multirow{2}{*}{\textbf{PCA AUPRC$^\star$}} & \multicolumn{4}{c}{\textbf{Deletion Performance}} \\ \cline{6-9}

& & & & & $\mathit{Total \ Deletion}$ & $\mathit{TPR^\star} \uparrow$ & $\mathit{FPR} \downarrow$ & $\mathit{Precision} \uparrow$ \\ \hline

\multirow{2}{*}{\textit{Enron Spam}} & EmbMarker & $0.9284$ & $0.9227$ & $0.9685$ & $572/5000$ & $91.49\%$ & $1.26\%$ & $90.21\%$ \\ \cline{2-2}

& WARDEN & $0.7348$ & $0.7348$ & $0.9530$ & $619/5000$ & $92.91\%$ & $2.14\%$ & $84.65\%$  \\ \hline

\multirow{2}{*}{\textit{SST2}} & EmbMarker & $0.8947$ & $0.8888$  &  $0.9214$ & $439/5000$  &  $95.68\%$ & $2.30\%$ & $75.63\%$   \\ \cline{2-2}

& WARDEN & $0.6190$ & $0.6190$ &  $0.9000$ & $437/5000$ & $95.68\%$ & $2.26\%$ & $75.97\%$  \\ \hline

\multirow{2}{*}{\textit{MIND}} & EmbMarker & $1.0$ & $1.0$ & $1.0$ & $152/5000$ & $100\%$ & 0\% &  $100\%$   \\ \cline{2-2}

& WARDEN & $0.4971$ & $0.4971$ & $0.7957$ & $188/5000$ & $84.21\%$ & $1.24\%$ & $68.09\%$   \\  \hline

\multirow{2}{*}{\textit{AG News}} & EmbMarker &  $0.5665$ & $0.5398$  & $0.7052$ & $1478/5000$ & $97.65\%$ & $19.62\%$ &  $42.08\%$  \\ \cline{2-2}

& WARDEN & $0.3323$ & $0.3323$ & $0.6791$  & $1498/5000$ & $96.86\%$ & $20.19\%$ & $41.19\%$   \\  

\noalign{\hrule height 0.3mm}

\end{tabular}


\end{threeparttable}

\end{minipage}

}

\caption{\label{table_7} Semantic Perturbation Attack Performance. `$\star$' demonstrates the most important metrics.}

\end{table*}






\begin{figure*}[!t]
    \centering
    \subfloat[Enron Spam]{\includegraphics[width=1.50in]{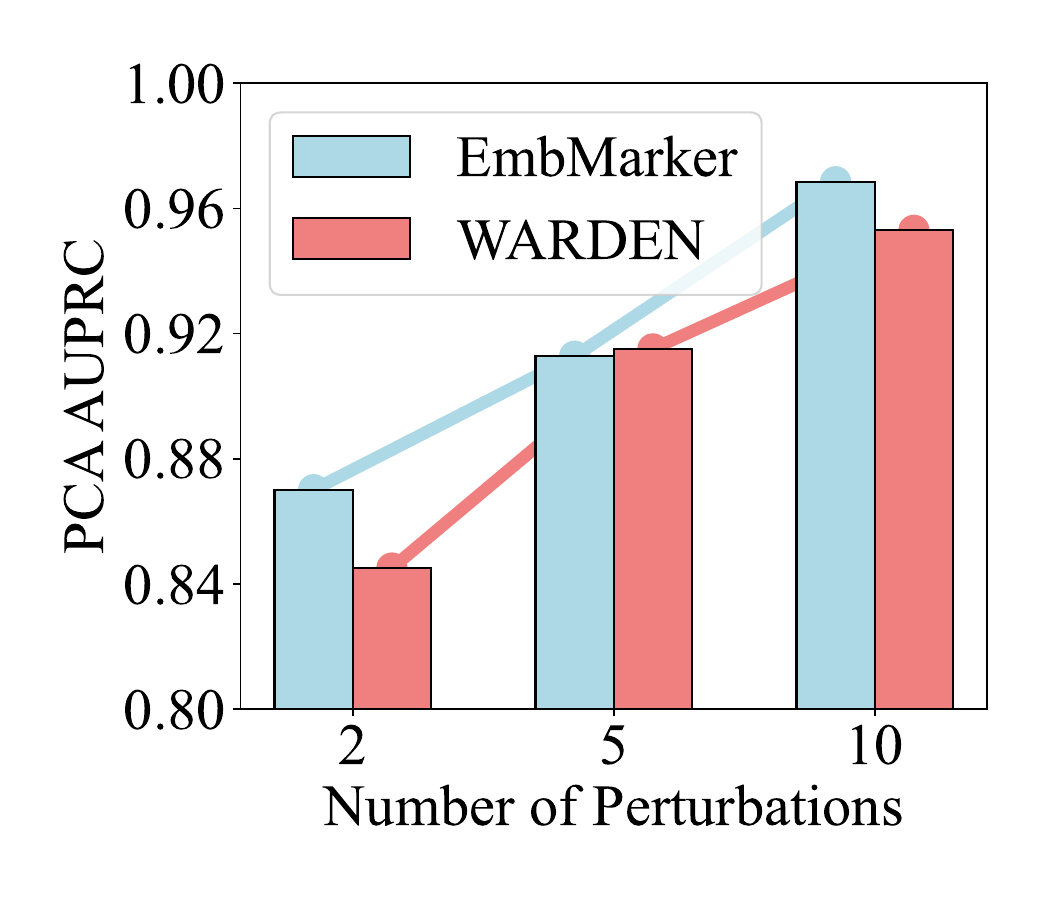}
    \label{ablation_1_enron}}
    \hspace{0.01cm}
    \subfloat[SST2]{\includegraphics[width=1.50in]{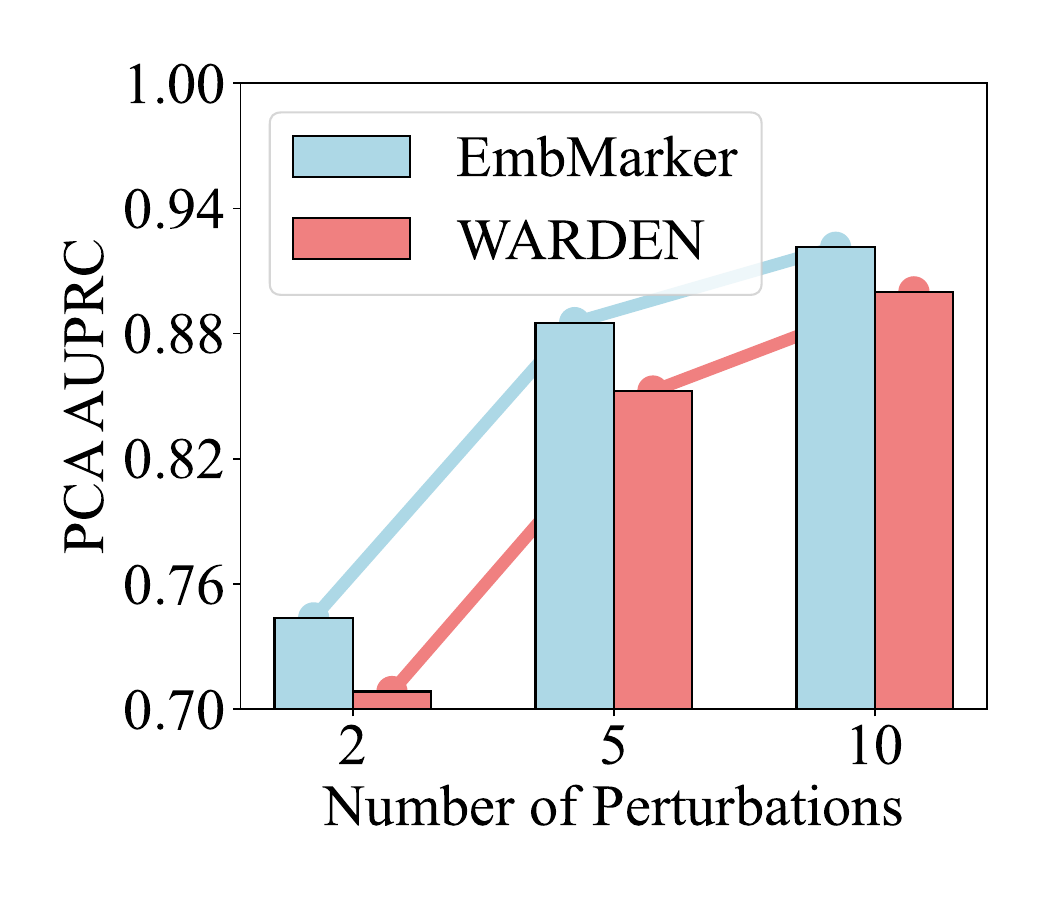}
    \label{ablation_1_sst2}}
    \hspace{0.01cm}
    \subfloat[MIND]{\includegraphics[width=1.50in]{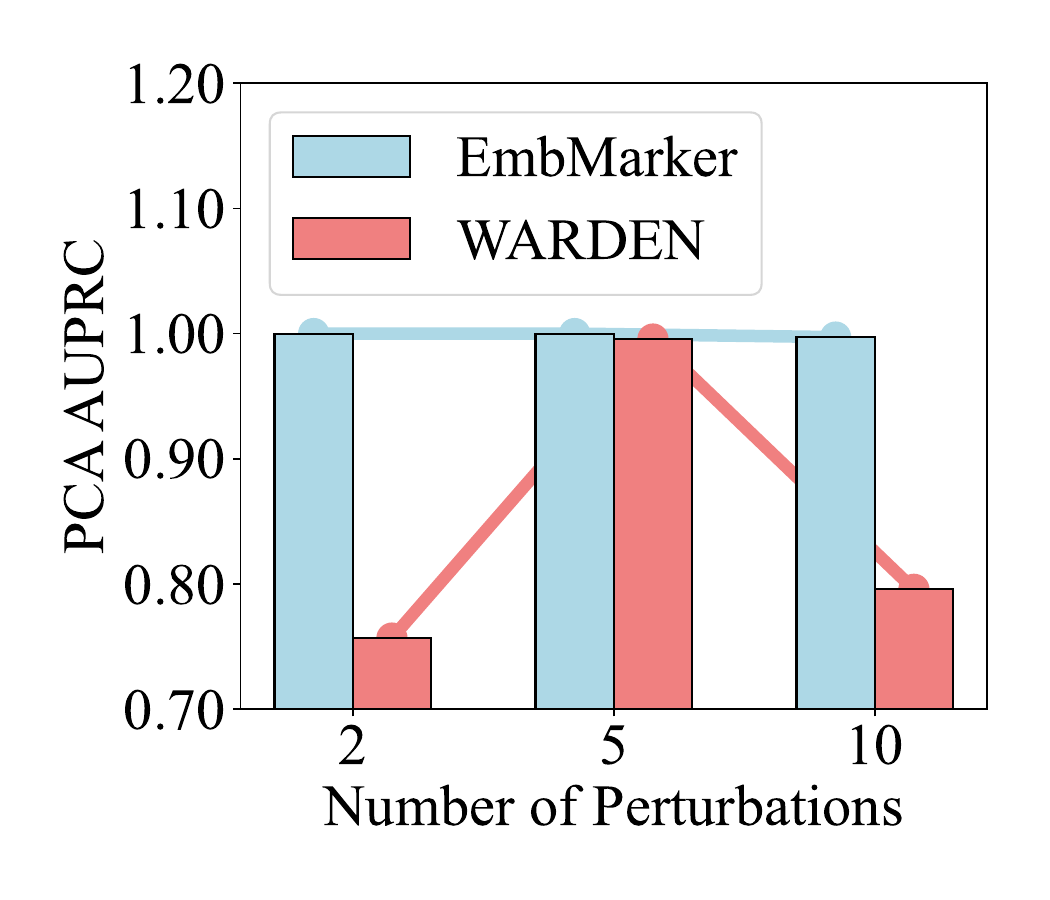}
    \label{ablation_1_mind}}
    \hspace{0.01cm}
    \subfloat[AG NEWS]{\includegraphics[width=1.50in]{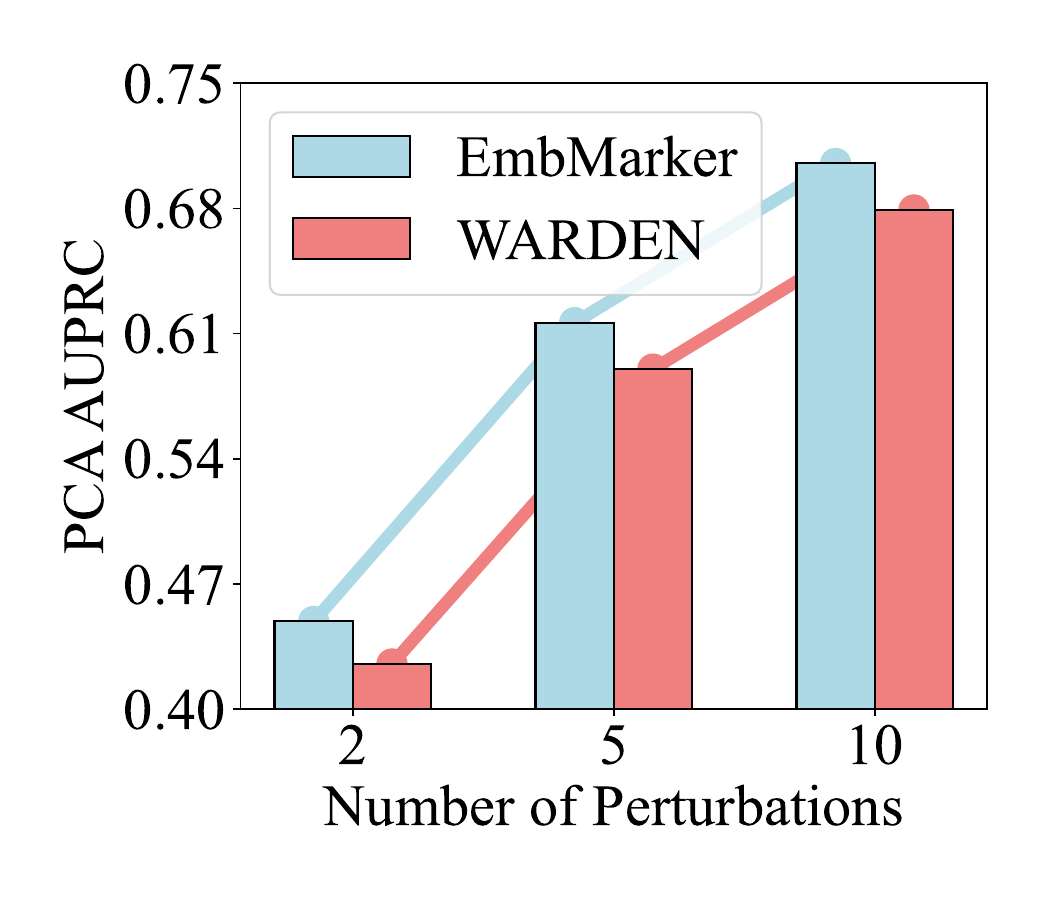}
    \label{ablation_1_ag}}
\caption{PCA AUPRC and Number of Perturbations. }
\label{ablation_1}
\end{figure*}


\begin{figure*}[!t]
    \centering
    \subfloat[Enron Spam]{\includegraphics[width=1.50in]{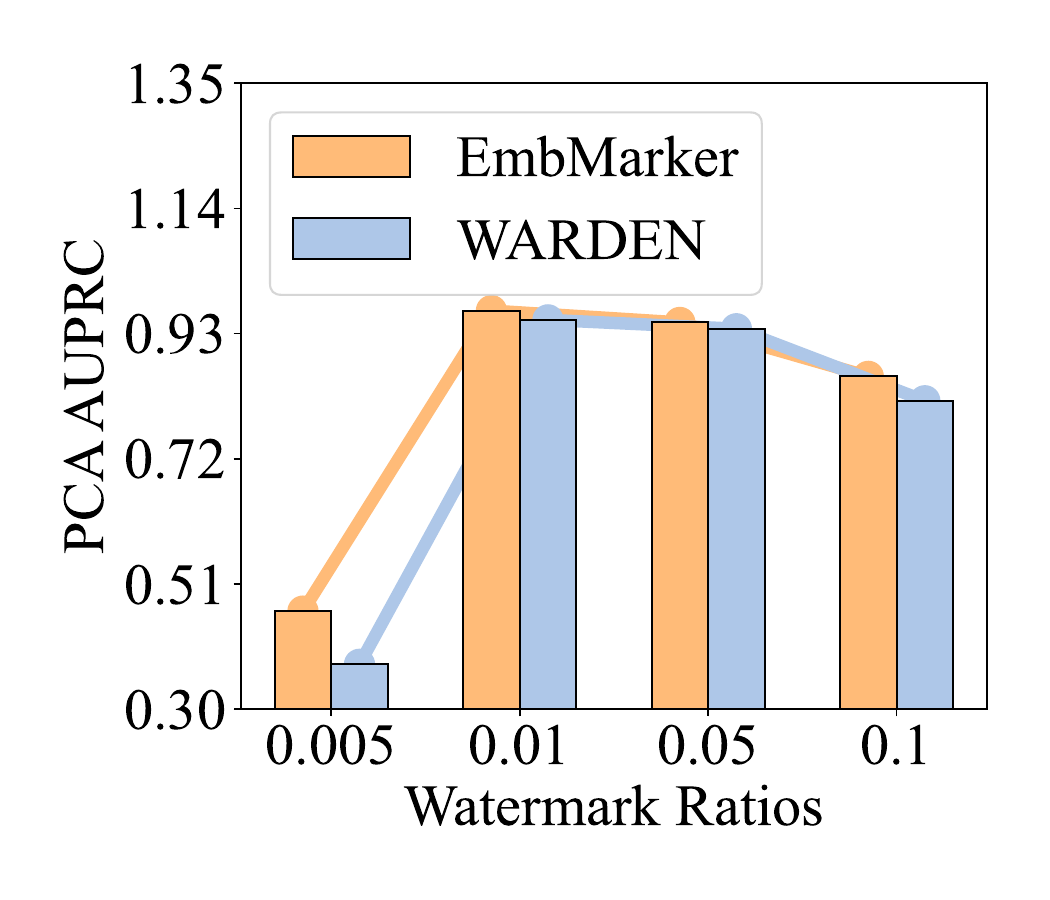}
    \label{ablation_2_enron}}
    \hspace{0.01cm}
    \subfloat[SST2]{\includegraphics[width=1.50in]{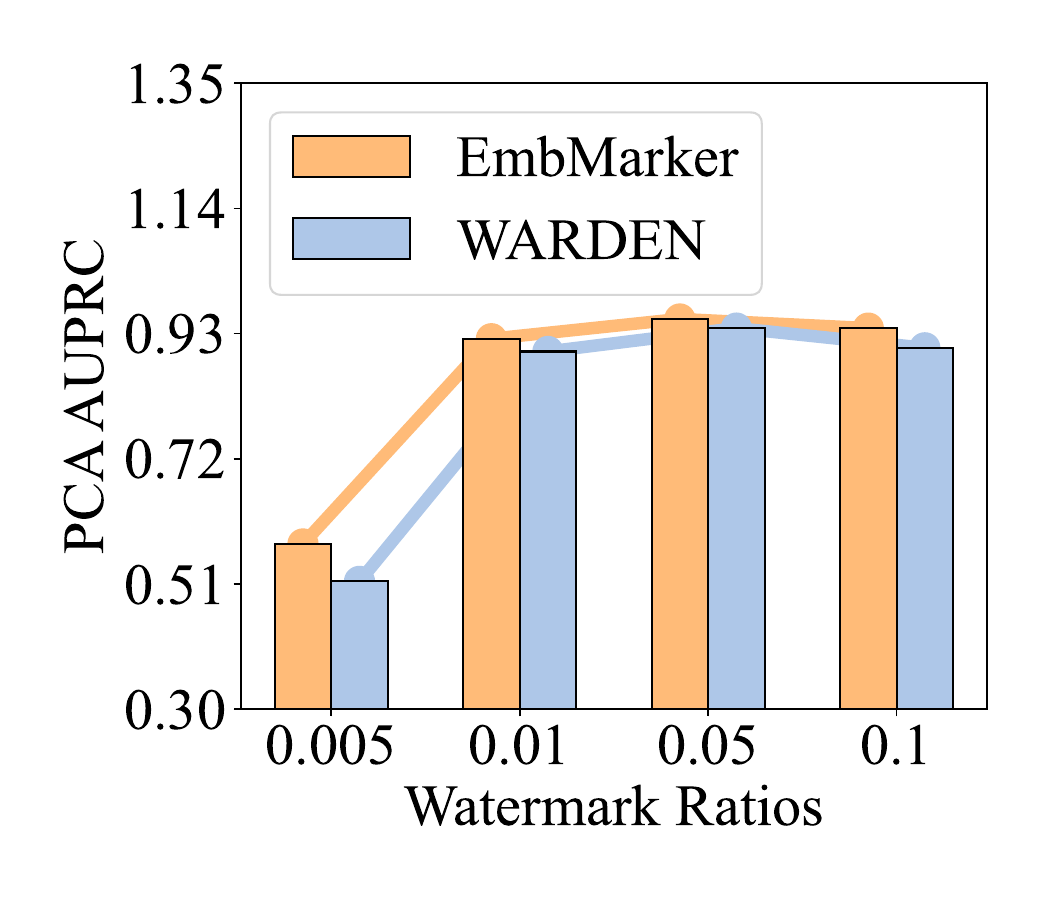}
    \label{ablation_2_sst2}}
    \hspace{0.01cm}
    \subfloat[MIND]{\includegraphics[width=1.50in]{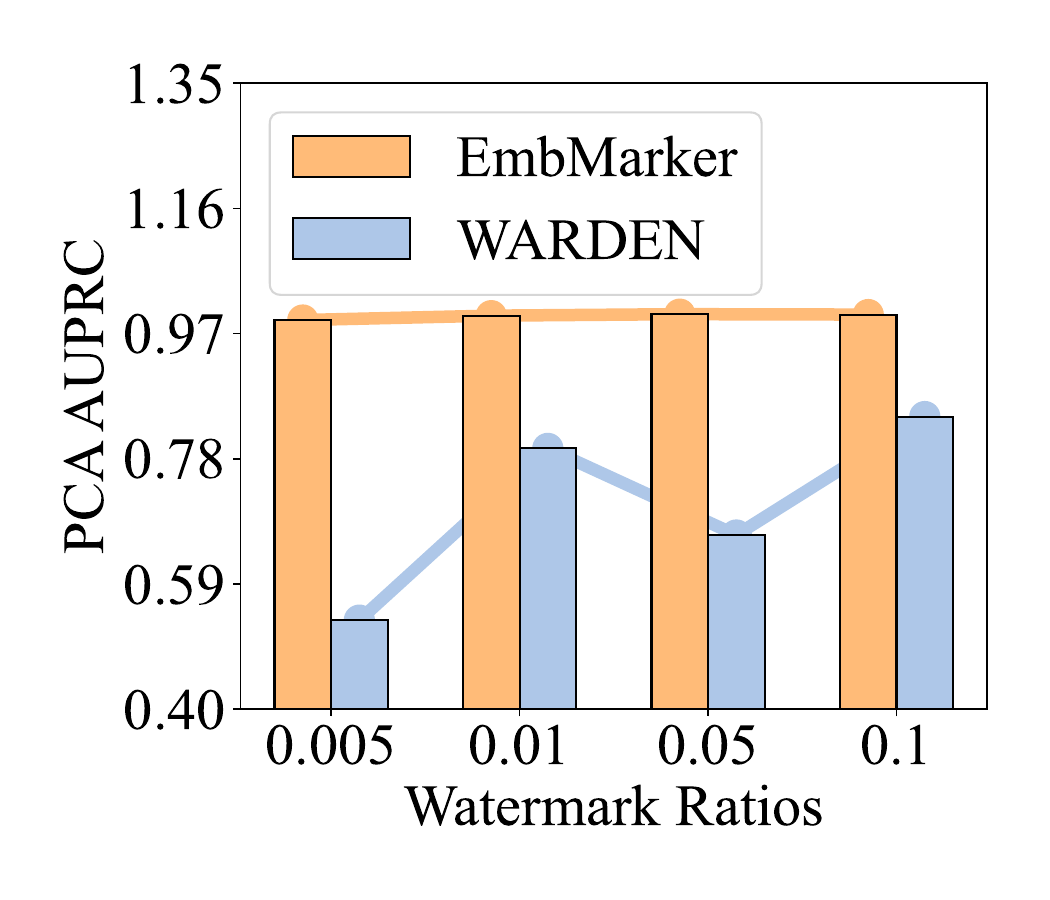}
    \label{ablation_2_mind}}
    \hspace{0.01cm}
    \subfloat[AG NEWS]{\includegraphics[width=1.50in]{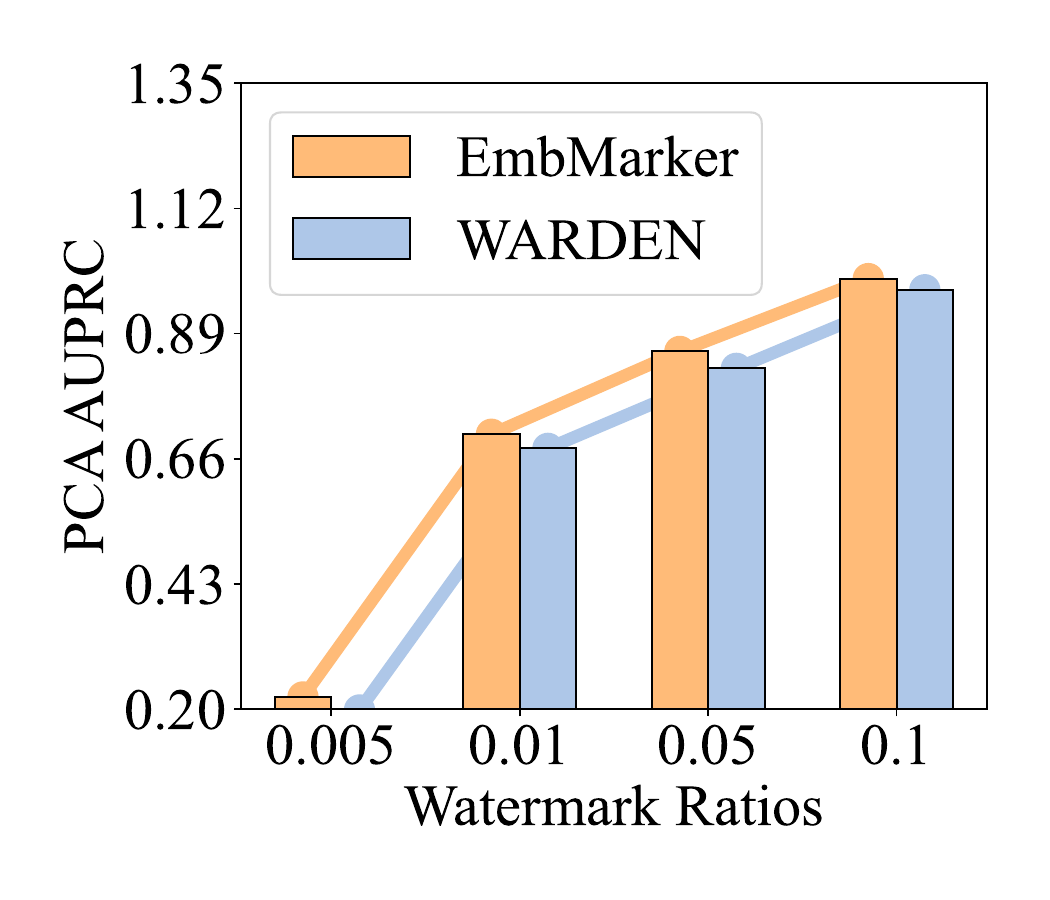}
    \label{ablation_2_ag}}
\caption{PCA AUPRC and Watermark Ratio.}
\label{ablation_2}
\end{figure*}

\subsection{Ablation Study}

\label{sec:ablation}

We conducted extensive experiments on SPA from multiple perspectives to validate its effectiveness and capability across various scenarios.


\textbf{\textit{PCA Score demonstrates superior robustness compared to other metrics.}} Table~\ref{table_7} shows that the PCA score metric remains stable across different schemes. Table~\ref{table_7} also shows the performance of watermark identification using the PCA score metric, along with a TPR universally exceeding 90\%. Furthermore, PCA score outperforms cosine similarity and $L_2$ distance, maintaining consistent better performance across schemes. This is likely because the PCA algorithm extracts and preserves the watermark information in the embeddings while eliminating redundant information.

\textbf{\textit{SPA performance improves as the number of semantic perturbations increases.}} We evaluated SPA performance under different numbers of perturbations using PCA AUPRC as the evaluation metric. The perturbation suffixes are selected following the order determined by suffix search guidance. The results shown in Figure \ref{ablation_1} indicate that, SPA performance increases and stabilizes as the number of perturbations grows. This further demonstrates the effectiveness of our attack strategy, as it ensures that effective suffixes are incorporated among multiple candidates.

\textbf{\textit{SPA remains effective under different watermark ratios.}} We evaluated SPA's performance under varying watermark ratios using PCA AUPRC, with a fixed number of perturbations. Figure \ref{ablation_2} shows that even with low watermark ratios (low-frequency triggers), SPA achieves a PCA AUPRC of 0.3-0.4, despite the stealer model failing to learn watermark behavior. Performance improves as the watermark ratio increases, though a slight AUPRC decline may occur when watermark ratio reaches 0.1. However, a high watermark ratio will result in excessive watermark injection and embedding modification. Nevertheless, the PCA AUPRC consistently remains above 0.9, demonstrating SPA's robustness across varying watermark ratios.


\section{Discussion of Mitigation Strategies}
\label{sec:discussion}

To counter SPA, we further explored potential mitigation strategies to address the effects of semantic perturbations. We suggest a deep learning-based solution with: (1) a semantic-aware injection model that dynamically embeds watermarks based on semantic features, and (2) a verification model. Integrating the adversarial noise module during training may improve resilience against virous attacks. The semantic-aware EaaS watermarking paradigm presents a promising SPA-resistant approach.



\section{Related Work}
\label{sec:related_work}

\subsection{Model Extraction Attack}

Model extraction attacks \cite{orekondy2019knockoff, sanyal2022towards, chandrasekaran2020exploring} threaten Deep Neural Networks (DNNs) and cloud services by enabling adversaries to replicate models without internal access. Attackers can query APIs \cite{kalpesh2020thieves} or gather physical data \cite{hu2020deepsniffer} to train the stolen models. Public APIs, especially in current EaaS services based on LLMs and MLLMs, are proved to be vulnerable \cite{liu2022stolenencoder, sha2023can}.

\subsection{Deep Watermarking}


Deep watermarking can be classified into white-box, black-box, and box-free approaches based on accessible data during verification \cite{li2021survey}. White-box watermarking schemes access model parameters \cite{yan2023rethinking, lv2023robustness, pegoraro2024deepeclipse}, while black-box schemes rely only on the model output \cite{leroux2024multi, Lv2024SSLWMAB}. Box-free watermarking schemes exploits inherent output variations without crafted queries \cite{an2024box}. In EaaS, watermarking can be regarded as a form of black-box watermarking.

\section{Conclusion}
\label{sec:conclusion}

In this paper, we propose SPA, a novel attack exploiting the limitation that current schemes rely solely on semantic-independent linear transformations. SPA conducts semantic perturbation to input text, constructs embedding pairs using the original and perturbed embeddings, and selectively deletes suspicious samples while preserving service utility. Our extensive experiments demonstrate the effectiveness of SPA. We also validate the importance of SPA's components and explore mitigation strategies. Our work emphasizes the critical role of text semantics in EaaS watermarking.


\section*{Limitations}
\label{sec:limitations}

In this paper, we propose SPA, a novel attack which exploits the semantic-independent vulnerabilities inherent in current EaaS watermarking schemes, successfully removing the majority of watermarked embeddings. However, an attacker requires a small local model for assistance to successfully execute SPA. Although such a scenario is realistic, we plan to explore attack schemes that do not require assistant models in our future work. Additionally, after each text perturbation, the attacker needs to re-access the original EaaS service, which increases the cost of SPA. Furthermore, we note that as the number of suffixes increases, the effectiveness of SPA becomes more stable, while an insufficient number of suffixes may lead to failure of SPA, thereby further amplifying concerns regarding the associated costs. In future, we believe that more advanced watermarking schemes will emerge, but SPA provides a perspective that emphasizes the importance of text semantics in the design of EaaS watermarking schemes. We will continue to explore how to develop more feasible attack and watermarking schemes with enhanced robustness.

\section*{Ethics Statement}
\label{sec:ethics}

We introduce a novel and effective attack targeting EaaS watermarks through the semantic perturbation. Our objective is to underscore the critical consideration of text semantics in EaaS watermark design, thereby enhancing security. We believe that the first step toward enhancing security is to expose potential vulnerabilities. All our experiments are conducted under control, with no attempts made to launch actual attacks on EaaS service providers. We have further explored potential mitigation strategies to address SPA.

\bibliography{custom}

\section*{Appendix} 

\appendix

\section{Overview of Different Attack Methods}

\label{appendix:attack_methods}

In Appendix \ref{appendix:attack_methods}, we provide a comprehensive and detailed introduction to various attack methods, including CSE, PA, and ESSA.

\begin{itemize}
    \item \textbf{CSE} \cite{shetty2024warden} is a kind of watermark elimination attack. CSE uses clustering to identify embedding pairs, selects potential watermarked embeddings by analyzing discrepancies between a standard model and the victim model, and eliminates principal components to erase watermark signals.

    \item \textbf{PA} \cite{shetty2024wet} is a kind of watermark elimination attack. PA employs a language model to rewrite input texts multiple times, retaining semantics but potentially losing trigger tokens. Averaging embeddings from these iterations dilutes the watermark signals. This attack paradigm modifies original embeddings, inevitably compromising the utility of embeddings.

    \item \textbf{ESSA} \cite{yang2024defending} is a kind of watermark identification attack. ESSA appends a token to the input text and evaluating whether the token functions as a trigger by analyzing the divergence between embeddings before and after token addition. 
\end{itemize}

\section{Definition of the Threat Model}

\label{appendix:threat_model}

In Appendix \ref{appendix:threat_model}, we clearly define the threat model, detailing the objective, knowledge, and capability of the attacker.

\textbf{Attacker's Objective.} TThe attacker aims to use embeddings from the victim model $\Theta_v$ without watermark verification. The attacker can then efficiently provide a competitive alternative instead of pre-training a new model.

\textbf{Attacker's Knowledge.} The EaaS service operates as a black box. The attacker queries the victim service $S_v$ using a dataset $D_c$, where each sample is $d_{c_i}$. While unaware any information of $\Theta_v$, the attacker can reasonably access a general text corpus $D_p$ and a small local embedding model $\Theta_s$ to design the attack algorithm.

\textbf{Attacker's Capability.} With sufficient budget, the attacker can query $S_v$ to obtain the embedding set $E_c$ for $D_c$. They can then employ various attack strategies to bypass watermark verification.

\section{Exploration of Perturbations}

\subsection{Exploration of Suffix}

\label{appendix:A}

\begin{figure}[!t]
    \centering
    \includegraphics[width=0.48\textwidth]{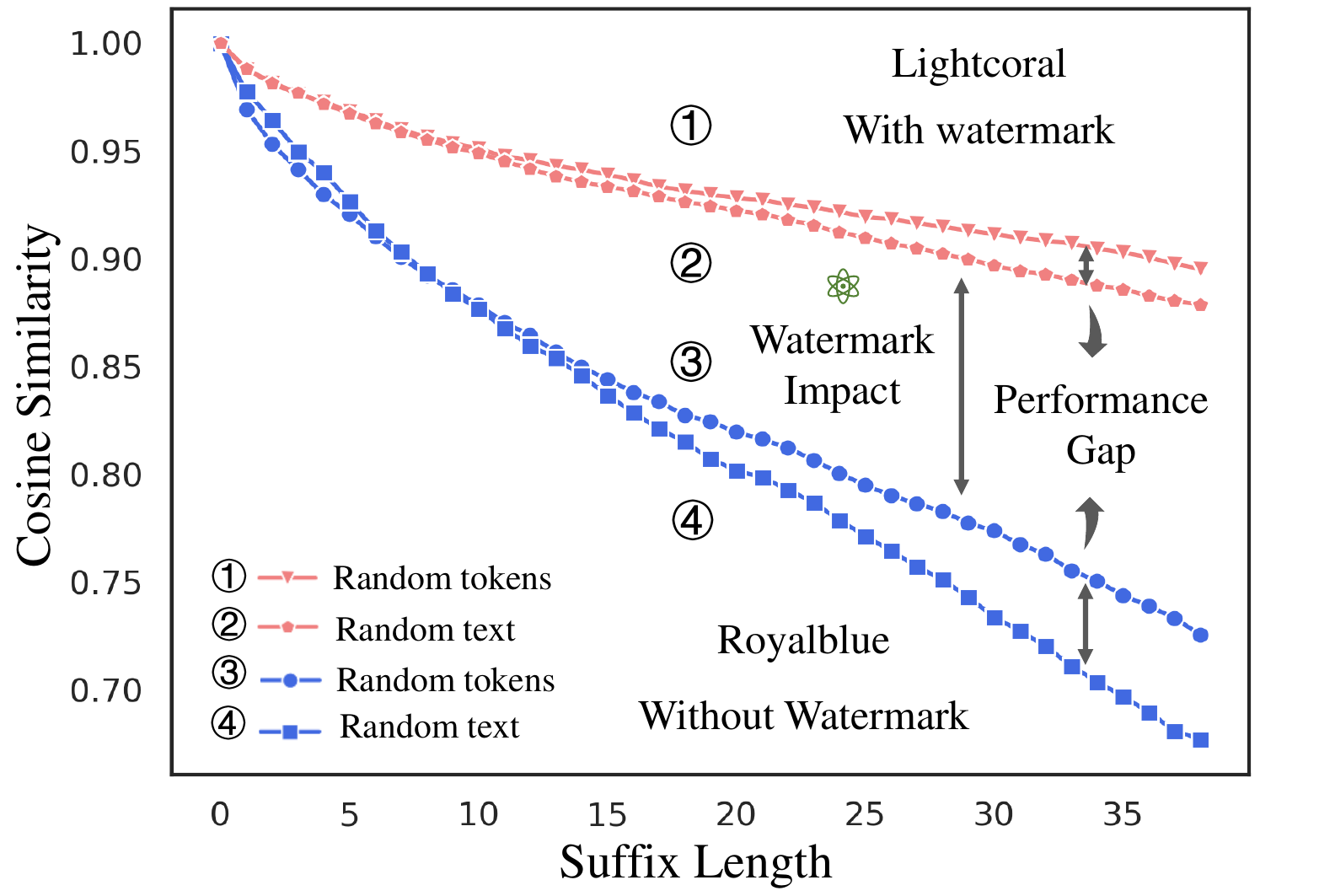}  
    \caption{Different Approaches of Semantic Perturbations: Length and Semantics. Regardless of whether watermarked or not, random text preforms better than random tokens. The injection of the watermark has led to a significant gap between the curves.}
\label{fig_4}
\end{figure}

In Appendix \ref{appendix:A}, we provide the detailed exploration of semantic perturbation. The text perturbation denoted as $perb$ can only be constructed as prefix or suffix. The potential construction space for the suffix can be classified from two perspectives: the length of the suffix and its semantics. We use EmbMarker \cite{peng2023you} as an example.

\textbf{\textit{Random tokens without semantics}:} We first explore a simple construction method by the adding random tokens as the suffix without semantics. Specifically, we tokenize each sentence in a general text corpus and compile all tokens into a total token vocabulary. We randomly add tokens to the suffix. At this stage, we explore the relationship between suffix length and perturbation performance before and after the watermark injection, measured by $(e_{c_i}, e'_{c_i})$. The results in Figure \ref{fig_4} indicate that as the suffix length increases, the embeddings similarity gradually decreases. After the watermark injection to $(e_{c_i}, e'_{c_i})$, the rate of decrease significantly slows and remains notably higher than the curve without the watermark injection.

\textbf{\textit{Random text with semantics}:} We randomly selected long texts from a general text corpus, tokenize it to obtain a sequence of tokens and sequentially add each token to the suffix. We explored the effects both with and without watermark injection. The results are illustrated in Figure \ref{fig_4}. It is evident that semantic suffix lead to a faster enhancement of perturbation performance, with the curve with watermark injection also significantly exceeding that without injection. Interestingly, for the same suffix length, the performance of perturbations using text with semantics is generally higher than that achieved with random tokens. The finding suggests that using the suffix with semantics is more cost-effective and produces better results. Therefore, we will consistently utilize the semantic suffix during the perturbation process.

\textbf{\textit{Text with \& without semantics}:} For suffix, the construction space can be categorized from two perspectives: length and semantics. A series of experiments demonstrate that using random text with semantics is more cost-effective and produces better results compared to random tokens without semantics. Based on this, we propose a heuristic perturbation scheme.

\subsection{Heuristic Perturbation Scheme}

\label{appendix:B}

\begin{figure}[htbp]
    \centering\includegraphics[width=0.48\textwidth]{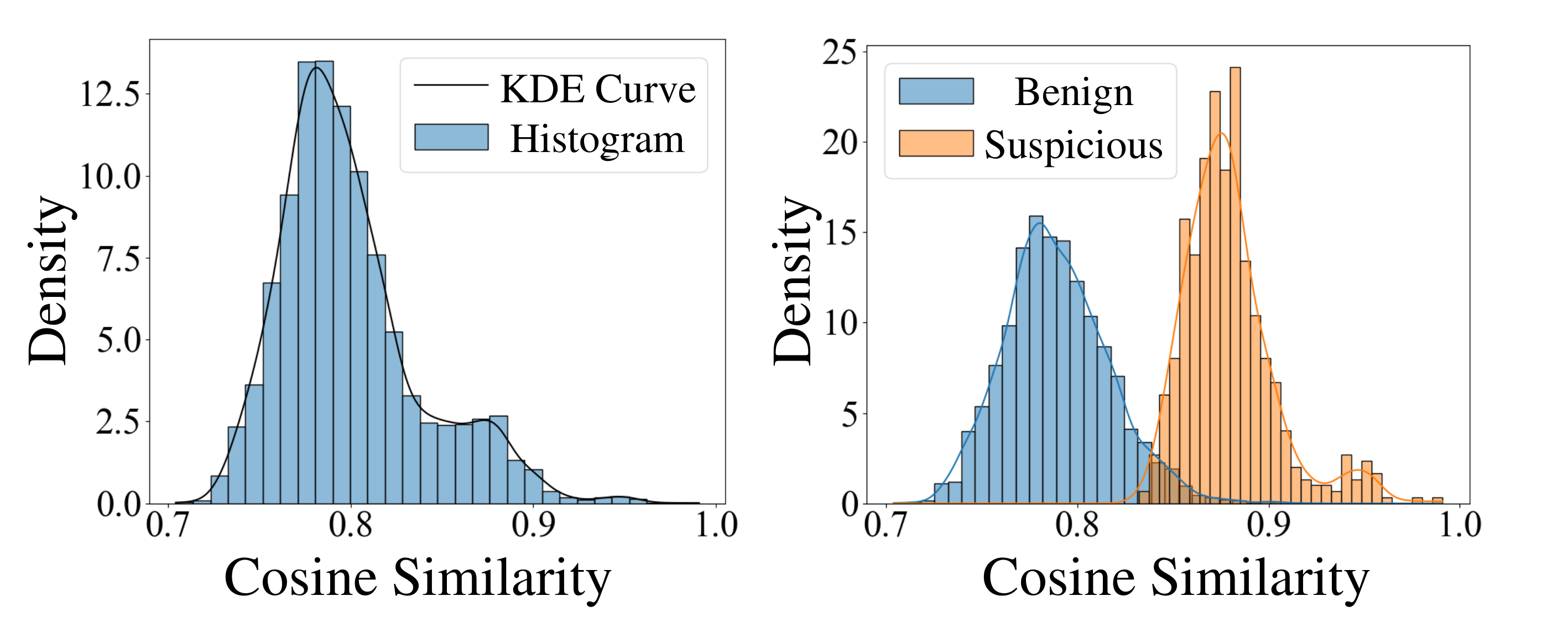}
    \caption{Cosine similarity metric distribution and KDE curve of the Enron Spam dataset in Heuristic Perturbation Scheme.}
\label{fig_5}
\end{figure}

In Appendix \ref{appendix:B}, we introduce heuristic semantic perturbation scheme. Semantic suffixes improve perturbation performance at lower costs, making suspicious samples easier to detect. Based on this, we propose a heuristic perturbation scheme. Following previous works, we focus on text classification tasks. In the context of text classification, heuristic perturbation scheme randomly selects samples with different labels from original as suffixes, leveraging semantic differences to enhance the perturbation. We randomly select $k$ samples for perturbation and calculate the average cosine similarity of $k$ embedding pairs, to reduce the influence of potential triggers in the suffixes. We conducted experiments on four classic datasets: Enron Spam \cite{metsis2006spam}, SST2 \cite{socher2013recursive}, MIND \cite{wu2020mind} and AG News \cite{zhang2015character}. From the perspectives of the attacker and ground truth, the cosine similarity distribution of Enron Spam dataset is shown in Figure \ref{fig_5}. The distribution results indicate observable differences for the Enron Spam and MIND datasets, while such differences are less pronounced for the SST2 and AG News datasets. Thus, we need to further explore a more effective approach.

\renewcommand{\arraystretch}{1.3}

\begin{table*}[htbp]

\centering

\scalebox{0.73}{

\begin{minipage}{1.25\textwidth}

\begin{threeparttable}
\begin{tabular}{c::c c c::c c c c c}
\noalign{\hrule height 0.3mm}

\textbf{Datasets} & \textbf{Train} & \textbf{Test} & \textbf{Class} & \textbf{Metrics} & \textbf{Schemes} & \textbf{Original} & \textbf{Subset}  &  \textbf{Epoch Adjustment}  \\ \hline

\multirow{2}{*}{\textit{Enron Spam}} & \multirow{2}{*}{$31,716 \rightarrow 5,000$} & \multirow{2}{*}{$2,000 \rightarrow 500$} & \multirow{2}{*}{$2$} & \multirow{2}{*}{ACC.(\%)} & EmbMarker &  $94.85\%$ & $92.00\%$ & $3 \rightarrow 20$  \\ \cline{6-6} 

& & & & & WARDEN & $94.60\%$ & $92.20\%$ & $3 \rightarrow 10$ \\ \hline

\multirow{2}{*}{\textit{SST2}} & \multirow{2}{*}{$67,349 \rightarrow 5,000$} & \multirow{2}{*}{$872 \rightarrow 500$} & \multirow{2}{*}{$2$} & \multirow{2}{*}{ACC.(\%)} & EmbMarker &  $93.46\%$ & $91.60\%$ & $3 \rightarrow 30$ \\ \cline{6-6}

& & & & & WARDEN & $93.46\%$ & $92.20\%$ & $3 \rightarrow 50$ \\ \hline

\multirow{2}{*}{\textit{MIND}} & \multirow{2}{*}{$97,791 \rightarrow 5,000$} & \multirow{2}{*}{$32,592 \rightarrow 500$} & \multirow{2}{*}{$18$} & \multirow{2}{*}{ACC.(\%)} & EmbMarker & $77.23\%$  & $69.20\%$ & $3 \rightarrow 75$ \\ \cline{6-6}

& & & & & WARDEN & $77.18\%$ & $71.80\%$ & $3 \rightarrow 75$ \\ \hline

\multirow{2}{*}{\textit{AG News}} & \multirow{2}{*}{$120,000 \rightarrow 5,000$} & \multirow{2}{*}{$7,600 \rightarrow 500$} & \multirow{2}{*}{$4$} & \multirow{2}{*}{ACC.(\%)} & EmbMarker & $93.57\%$  & $88.80\%$ & $3 \rightarrow 20$  \\ \cline{6-6}

& & & & & WARDEN & $93.76\%$  & $89.00\%$ & $3 \rightarrow 20$ \\ 
\noalign{\hrule height 0.3mm}

\end{tabular}

\end{threeparttable}

\end{minipage}

}

\caption{\label{table_3} Training Settings.}

\end{table*}

\subsection{Semantic Perturbation Guidance}

\label{appendix:C}

\begin{algorithm}[!t]
\caption{Suffix Perturbation Guidance}
\small
\begin{algorithmic}[1]
\STATE \textbf{Input:} Perturbation Pool $P$, Dataset $D_c$
\STATE \ \ \ \ \ \ \ \ \ \ \ \ Standard Model $\Theta_s$, Hyperparameter $k$
\STATE \textbf{Output:} Metric Values $v$
\STATE Initialize $s \gets \emptyset$$($Suffix$)$
\STATE Initialize $n \gets |D_c|$, $m \gets |P|$
\STATE Set $max(s) \gets 1$ \hfill \COMMENT{$\triangleright$ Cosine similarity range: [-1, 1]}

\FOR{$i = 1$ to $n$}
    \FOR{$j = 1$ to $m$} 
        \STATE $d'_{c_i} \gets d_{c_i}+perb_j$
        \STATE Encode: 
        $se_{c_i} \gets \Theta_s(d_{c_i})$, 
        $se'_{c_i} \gets \Theta_s(d'_{c_i})$
        \STATE $sim \gets \textit{cosine}(se_{c_i}, se_{perb})$ 
        \IF{$|s| < k$}
            \STATE Append $perb_j$ to $s$
        \ELSIF{$|s| \geq k$ \AND $sim < max(s)$}
            \STATE Remove $max(s)$ from $s$
            \STATE Insert $perb_j$ into $s$
        \ELSE
            \STATE Skip $perb_j$
        \ENDIF
    \ENDFOR
    \STATE Compute aggregate metric: 
    $metric \gets agg(s)$
    \STATE Append $metric$ to $v$
\ENDFOR
\RETURN $v$
\end{algorithmic}
\label{algorithm_1}
\end{algorithm}

In Appendix \ref{appendix:C}, we introduce another small local model suffix perturbation guidance approach. The results in Figure \ref{fig_5} indicate that the effectiveness of the simple heuristic perturbation scheme needs further improvement. Although the embedding spaces of $\Theta_v$ and $\Theta_s$ differ, the variations between $(e_{c_i}, e'_{c_i})$ under the same perturbation show similar patterns across all these spaces. Specifically, we input the text pair $(d_{c_i}, d_{c_i}+perb)$ into $\Theta_s$ to obtain the corresponding embedding pair $(se_{c_i}, se'_{c_i})$. The perturbation $perb$ traverses through all candidates in the perturbation pool. The $top$-$k$ $perb$ texts that minimize the similarity of $(se_{c_i}, se'_{c_i})$ are selected as candidate suffixes. Since the embeddings output by $\Theta_s$ are not watermarked, it is feasible to use this small local model to guide the perturbations for $\Theta_v$. We similarly take the aggregate metric over $k$ perturbed samples for evaluation. $\Theta_s$ captures the differential features between $(d_{c_i}, d_{c_i}+perb)$. Such differential features are consistent across models. However, suffix perturbation guidance is less efficient since each text have to traverse all the candidates in the perturbation pool. It results in the time complexity of $|D_c| \cdot |perb \ pool|$, requiring $\Theta_s$ to encode $|D_c| \cdot |perb \ pool|$ perturbation processes. The entire process of the algorithm is shown in Algorithm \ref{algorithm_1}.

\section{Dataset Introduction}

\label{appendix:dataset}

In Appendix \ref{appendix:dataset}, we will provide a comprehensive description of the specific details of the datasets utilized, including their structure, preprocessing steps, and relevant statistics. The datasets selected for our experiments—Enron Spam \cite{metsis2006spam}, SST2 \cite{socher2013recursive}, MIND \cite{wu2020mind}, and AG News \cite{zhang2015character}—are widely recognized as benchmark datasets in the field of Natural Language Processing (NLP). We apply the four datasets to the text classification task, with a primary focus on investigating the potential impact of watermarks on this downstream task.

\begin{itemize}
    \item \textbf{Enron Spam:} The Enron Spam dataset consists of the emails collection labeled as either ``spam" or ``non-spam" (ham), making it a valuable resource for studying spam filtering, email classification.
    \item \textbf{SST2:} The SST2 dataset is a collection of movie reviews labeled with binary sentiment (positive or negative), commonly used for training and evaluating models in sentiment classification tasks.
    \item \textbf{MIND:} The MIND dataset is a large-scale dataset designed for news recommendation. It can also used for news classification tasks.
    \item \textbf{AG News:} The AG News dataset is a collection of news articles categorized into four topics, commonly used for text classification and NLP tasks.
\end{itemize}

\section{Experiment Settings}

\label{appendix:D}

In Appendix \ref{appendix:D}, we will provide a detailed description of the training configurations employed in our experiments. Furthermore, we demonstrate that our experimental setup is both rational and effective in conducting various evaluation tests.

Table~\ref{table_3} provides detailed information about the datasets used in our study. It also highlights the adjustments made to the number of training epochs in order to ensure performance on the respective subsets of each dataset. Specifically, the smallest dataset contains more than 30,000 data items, while the largest dataset includes over 12,000 data items. For our experiments, we sampled a subset of 5,000 examples from the training set and 500 examples from the test set. This sampling strategy was carefully chosen to balance the need for the cost of the experiment with the goal of maintaining representative data coverage. Table~\ref{table_3} indicates that, despite using subsets, the accuracy of downstream tasks has not significantly decreased in different watermarking schemes. On certain specific datasets, the accuracy achieved using the subset for training has even shown a slight improvement. This may be attributed to the inherent randomness in training process. Since the focus is on a relatively simple text classification task, the model appears to perform well even on the subset, maintaining favorable results. The results of the experiments demonstrate that conducting tests on these subsets not only produces valid and meaningful outcomes but also confirms the practicality.

\end{document}